\documentclass[a4paper,doublecolumn]{IEEEtran}
\def\BibTeX{{\rm B\kern-.05em{\sc i\kern-.025em b}\kern-.08em
    T\kern-.1667em\lower.7ex\hbox{E}\kern-.125emX}}
\usepackage{booktabs}
\usepackage{epsfig,epstopdf,graphicx,psfrag,amsmath,cases}
\usepackage{latexsym,amssymb,amsmath,epsfig,algorithm,amsthm}
\usepackage{algorithmic}
\usepackage{nopageno}
\usepackage{color}
\usepackage{url}
\usepackage{amsthm}
\usepackage{scrtime}
\usepackage{bm}
\usepackage{cancel}
\usepackage{graphicx}
\usepackage{hyperref}
\usepackage{bbding}
\usepackage{multicol}
\usepackage{graphicx}
\usepackage{amsthm}
\usepackage{cite}
\usepackage{array}
\usepackage{float}
\usepackage[utf8]{inputenc}
\usepackage[english]{babel}
\usepackage{setspace}
\usepackage{subcaption}
\addto\captionsenglish{}
\usepackage[font=small]{caption}
\usepackage{pst-node}
\usepackage{mathtools,graphicx}

\begin{document}
\title{Beamforming Design for Intelligent Reflecting Surface-Enhanced Symbiotic Radio Systems}


\author{\IEEEauthorblockN{Shaokang Hu$^*$, Chang Liu$^*$, Zhiqiang Wei$^{\S}$, Yuanxin Cai$^*$, Derrick Wing Kwan Ng$^*$, and Jinhong Yuan$^*$ \\
          $^*$School of Electrical Engineering and Telecommunications, University of New South Wales, Sydney, Australia\\ $^{\S}$The Institute for Digital Communications (IDC), Friedrich-Alexander University Erlangen-Nuremberg, Germany\vspace{-9mm}}}

{\newtheorem{Thm}{Theorem}}
\newtheorem{theorem}{Theorem}
\newtheorem{Lem}{Lemma}
\newtheorem{Cor}{Corollory}
\newtheorem{Def}{Definition}
\newtheorem{Exam}{Example}
\newtheorem{Alg}{Algorithm}
\newtheorem{Prob}{Problem}
\newtheorem{Proof}{Proof}
\newtheorem{Remark}{Remark}
\newcommand{\abs}[1]{\lvert#1\rvert}
\newcommand{\norm}[1]{\lVert#1\rVert}


\maketitle

%
\thispagestyle{empty}
\begin{abstract}
This paper investigates multiuser multi-input single-output downlink symbiotic radio communication systems assisted by an intelligent reflecting surface (IRS). Different from existing methods ideally assuming the secondary user (SU) can jointly decode information symbols from both the access point (AP) and the IRS via multiuser detection, we consider a more practical SU that only non-coherent detection is available.
To characterize the non-coherent decoding performance, a practical upper bound of the average symbol error rate (SER) is derived.
Subsequently, we jointly optimize the beamformer at the AP and the phase shifts at the IRS to maximize the average sum-rate of the primary system taking into account the maximum tolerable SER constraint for the SU. To circumvent the couplings of variables, we exploit the Schur complement that facilitates the design of a suboptimal beamforming algorithm based on successive convex approximation. Our simulation results show that compared with various benchmark algorithms, the proposed scheme significantly improves the average sum-rate of the primary system, while guaranteeing the decoding performance of the secondary system.
\end{abstract}
\vspace{-5mm}
\section{Introduction}\vspace{-0mm}
Recently, intelligent reflecting surface (IRS)-assisted symbiotic radio (SR) systems \cite{long2019symbiotic} have been proposed as one of the promising technologies to achieve spectrally and energy-efficient transmission towards the sixth-generation (6G) communications. By exploiting the IRS, SR systems not only enhance the quality of the primary transmission from the access point (AP) to its  primary users (PUs), but also allow the IRS to be served as a secondary transmitter to convey its information to the desired secondary users (SUs).
Conventionally, an IRS consists of a large number of low-cost passive reflection elements (REs) and the phase shift of each element can reflect/redirect the incident signals to the desired users in a nearly-passive manner. Thus, the introduction of an IRS can further enhance the quality of primary transmission by providing a controllable additional signal propagation path for dedicated energy-focusing and energy-nulling \cite{wu2019intelligent}. On the other hand, in an SR system, an environment sensor serving as an information source can be connected to the IRS \cite{yan2019passive} for collecting the environmental information such as light intensity, temperature, and humidity. Thus, the IRS has its need to transmit the sensed information to a low data-rate SU. In these scenarios, the IRS can embed  its information to the reflected radio frequency signals originating from the AP. As such, a mutualistic SR system can be established by intelligent synergistic resources exchanges. Specifically, the SU shares the same frequency spectrum, energy, and infrastructure with PUs, which results in more spectrally and energy-efficient communications compared with conventional networks \cite{kammoun2020asymptotic,dong2020secure}.

To realize practical IRS-assisted SR systems, various schemes have been proposed. For instance, in \cite{hua2021novel,zhang2021reconfigurable}, the IRS is able to transmit information to the SU by adopting binary phase shift keying modulation to modulate its information over the incident signals from the AP. Yet, having the off state of all IRS elements concurrently deflects the purpose of deploying an IRS as it leads to a low spectral efficiency in end-to-end information transmission in the primary system.
 To further improve the spectral efficiency of IRS-assisted systems, \cite{yan2019passive} and \cite{lin2020reconfigurable} adopted a higher order IRS modulation by exploiting spatial modulation over IRS elements. However, the problem formulations in \cite{yan2019passive} and \cite{lin2020reconfigurable} did not take into account the quality-of-service (QoS) requirement of decoding IRS symbols at the SU. As such, the performance of the SU cannot be guaranteed. Moreover, all the aforementioned papers, i.e., \cite{hua2021novel,yan2019passive,zhang2021reconfigurable,lin2020reconfigurable}, ideally assumed that the SU is capable to perform sophisticated multiuser detection or successive interference cancellation for decoding the information symbols of the AP and the IRS jointly, which is generally impossible for a low-cost SU. Besides, without knowing the symbols transmitted by the AP, the effective channel state information (CSI) is generally unknown at the desired SU as it is a product of instantaneous CSI and AP symbols. As a result, it is challenging, if not impossible, for the implementation of the coherent detection proposed in \cite{hua2021novel,zhang2021reconfigurable}.
 Thus, a more practical resource allocation design for non-coherent detection in SR systems is desired.

In this paper, we consider an IRS-assisted multiuser multi-input single-output (MISO) downlink multiuser SR system, where the IRS can transmit its information to the SU while assisting the primary transmission between the AP and PUs. In particular, the IRS can modulate its information by applying an on/off  multi-level amplitude modulation to the index of the IRS elements.
In contrast to existing methods, e.g. \cite{hua2021novel,yan2019passive,zhang2021reconfigurable,lin2020reconfigurable}, we investigate a practical SR system which is able to acquire modulated IRS symbols at the SU without decoding AP symbols. In particular, non-coherent detection is adopted at the SU for decoding the IRS symbols. To quantify the decoding performance of the SU, we first derive an upper bound of the average symbol error rate (SER).
Furthermore, by jointly designing the precoding vector at the AP and the phase shift matrix at the IRS, the average sum-rate of the primary system is maximized subject to an SER-based constraint for the SU. Due to the coupling among the optimization variables and the explicit expression of the derived SER upper bound, the formulated problem is non-convex such that obtaining an optimal solution in polynomial time is generally intractable. As a compromise, we exploit the Schur complement and adopt successive convex approximation (SCA) to obtain a
suboptimal solution of the beamforming design problem. Simulation results show that the proposed scheme not only guarantees the decoding performance of the secondary system, but also significantly improves the average sum-rate of the primary system compared with various benchmarks.
\begin{figure}[t] 
  \centering
  \includegraphics[width=3in]{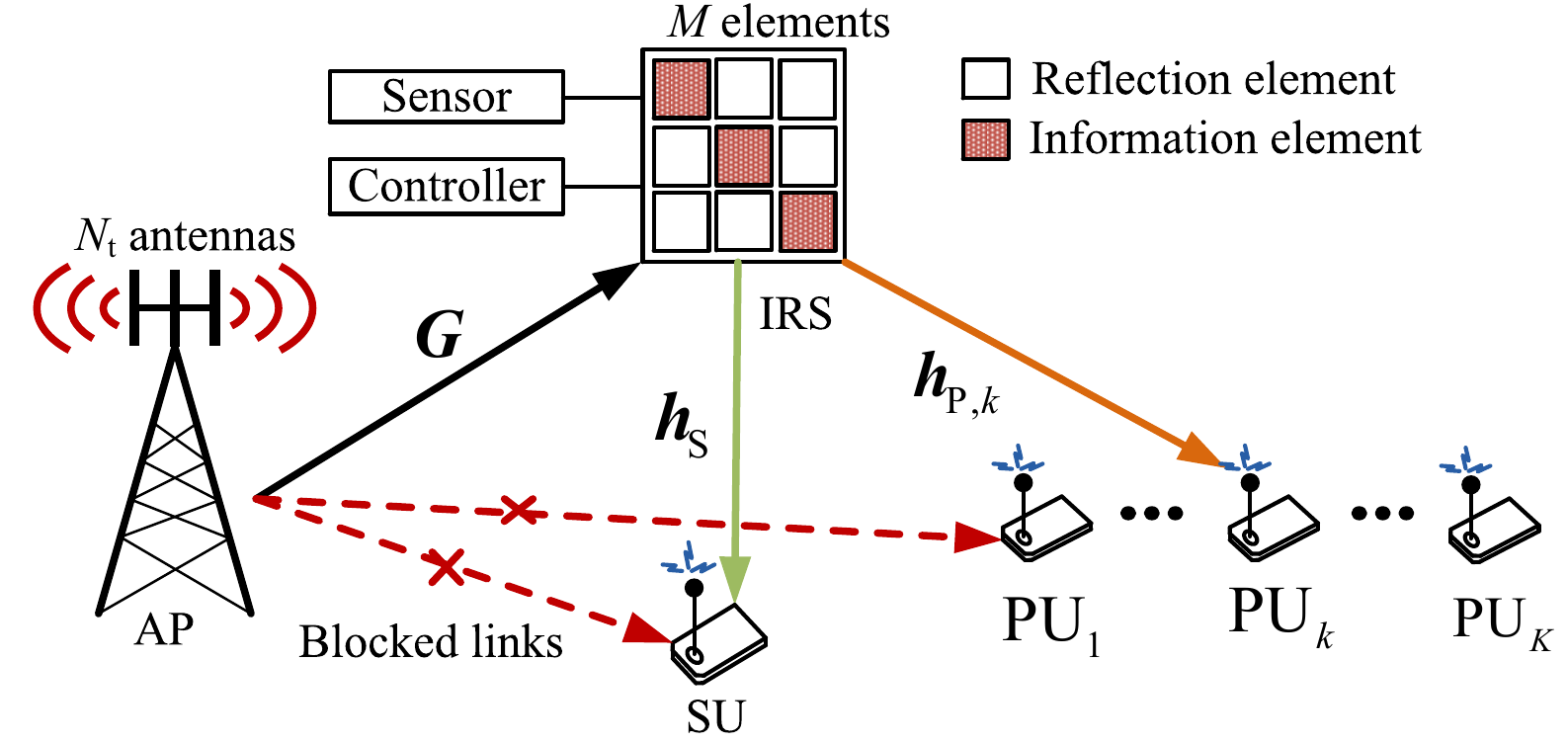} 
  \caption{\hspace{-1mm} An IRS-assisted SR system model.}
  \label{Fig:system_model} \vspace*{-6mm}
\end{figure}

\emph{Notations}: Scalars, vectors, and matrices are represented by lowercase letter $x$, boldface lowercase letter $\mathbf{x}$, and boldface uppercase letter $\mathbf{X}$, respectively. $\mathbb{B}^{N \times M}$ and $\mathbb{C}^{N \times M}$ denote the spaces of $N \times M$ matrices with binary and complex entries, respectively. $\mathbf{X}(n,m)$ denotes the element at the $n$-th row and the $m$-column of the matrix. 
The Euclidean norm and Frobenius norm of a vector/matrix are denoted by $\|\cdot\|$ and $\|\cdot\|_{\mathrm{F}}$, respectively. The absolute value of a complex-valued
scalar is denoted by $|\cdot|$. The occurrence probability of an event is denoted by $\mathrm{Pr}\{\cdot\}$. The conditional probability density function of $x$ on event $\mathcal{H}$ is denoted by $p(x|\mathcal{H})$. The transpose, conjugate transpose, conjugate, expectation, and trace of a matrix/vector are denoted by  $(\cdot)^{\mathrm{T}}$, $(\cdot)^{\mathrm{H}}$, $(\cdot)^{*}$, $\mathbb{E}[\cdot]$, and $\mathrm{Tr(\cdot)}$, respectively.
$\mathbf{X}\succeq\mathbf{0}$ and $\mathbf{X}\preceq\mathbf{0}$ mean that matrix $\mathbf{X}$ is positive semi-definite and negative semi-definite, respectively.
$\mathrm{diag(\mathbf{x})}$ denotes a diagonal matrix with its diagonal elements given by vector $\mathbf{x}$.
$j$ denotes the imaginary unit.
The distribution of a circularly symmetric complex Gaussian (CSCG) random variable with mean $\mu$ and variance $\sigma^2$ is denoted by $\mathcal{CN}(\mu, \sigma^2)$ and $\sim$ stands for ``distributed as''. The distribution of Erlang distribution is denoted by $\mathrm{Erlang}(\alpha,n)$ with scale parameter $\alpha$ and shape parameter $n$. $\mathbf{I}_N$ denotes an $N\times N$ identity matrix.
\vspace*{-4mm}
\section{System Model}\vspace{-1mm}
As shown in Fig. \ref{Fig:system_model}, we consider an IRS-assisted SR system, which includes an AP equipped with $N_{\mathrm{t}}>1$ antennas, an IRS with $M>1$ elements, $K>1$ single-antenna PUs, and a single-antenna SU\footnote{The extension to the case of multiple SUs will be considered in our future work.}.
In particular, the AP transmits $K$ independent data streams to $K$ PUs simultaneously with the assistance from the IRS. The precoding vector for the $k$-th PU ($\mathrm{PU}$$_k$) adopted at the AP is defined as $\mathbf{w}_k\in \mathbb{C}^{N_{\mathrm{t}}\times 1}, \forall k\in\mathcal{K}=\{1,\ldots,K\}$. Meanwhile, the IRS passively transmits the sensed environmental information of the connected sensor to the SU by altering the IRS reflection patterns via index modulation, which will be detailed in Sections II-B and II-C.
  The IRS reflection matrix is defined as $\mathbf{\Phi} =\mathrm{diag}($$e^{j\theta_1},\ldots,  e^{j\theta_m}$$ ,\ldots, e^{j\theta_M}) $$\in \mathbb{C}^{M\times M}$ with $\theta_m \in [0,2\pi), \forall m\in\mathcal{M}=\{1,\ldots,M\}$, denoting the phase shift at the $m$-th IRS
  element. For the ease of practical implementation of the IRS, the amplitude coefficients of all the elements are fixed to be unity in this paper. 
 Besides, as depicted in Fig. \ref{Fig:system_model}, the direct links of AP-to-PUs and AP-to-SU are blocked due to heavy shadowing. The aforementioned assumptions are commonly adopted in the literature \cite{wu2019intelligent,kammoun2020asymptotic,dong2020secure}. On the other hand, this paper considers a quasi-static flat fading channel model. We assume that handshaking has been performed between the AP, PUs, and the SU at the beginning of transmission. As such, the  channel coefficients of the AP-to-IRS link,  the IRS-to-PU$_k$  link,  and the IRS-to-SU  link can be acquired by exploiting some existing advanced channel estimation methods, e.g. \cite{liu2020deep}, which are denoted by  $\mathbf{G}\!\in\!\mathbb{C}^{M\!\times\! N_{\mathrm{t}}}$, $\mathbf{h}_{\mathrm{P},k}\!\in\!\mathbb{C}^{M\!\times \!1}$, and $\mathbf{h}_{\mathrm{S}}\!\in\!\mathbb{C}^{M\!\times \! 1}$, respectively.
\vspace*{-3mm}
\subsection{Transmission Framework}\vspace*{-1mm}
\begin{figure}[t] 
  \centering
  \includegraphics[width=3in]{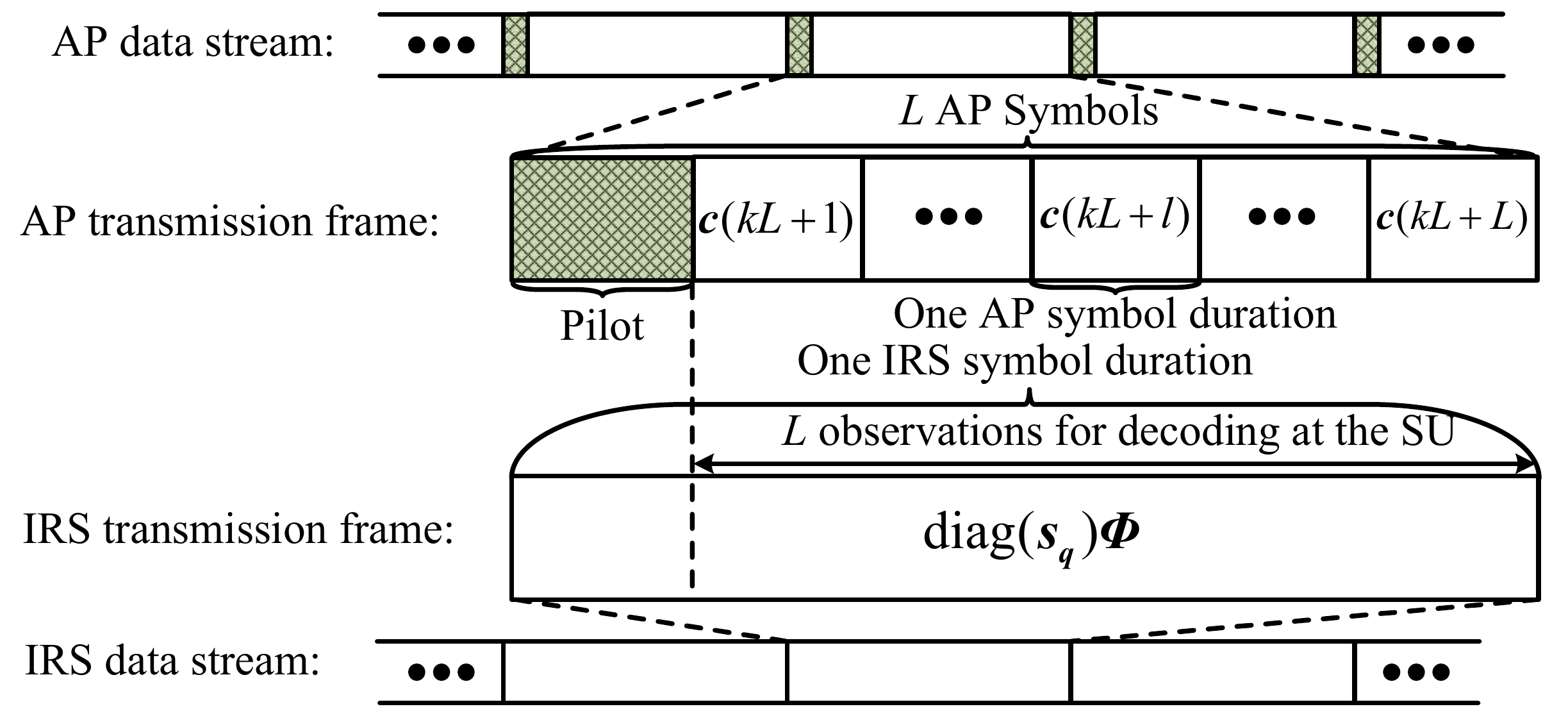} 
  \caption{\!\! A transmission frame structure of the considered IRS-assisted SR network. The AP and the IRS
  operate concurrently such that $L$ AP symbols duration is equivalent to that of $1$ IRS symbol.}
  \label{Fig:transmissionFrame} \vspace*{-6mm}
\end{figure}
Fig. \ref{Fig:transmissionFrame} shows the transmission frame structure of the considered IRS-assisted SR network, where the AP and the IRS simultaneously transmit their information symbols.
In particular, $\mathbf{s}_q\in  \mathbb{B}^{M\!\times \!1}$ denotes the IRS symbol to be transmited in each IRS transmission frame, which spans over a pilot sequence and $L$ AP symbols, $c_k(l)\sim\mathcal{CN}(0,1),\forall k, l \!\in\!\mathcal{L}\!=\!\{1,...,L\}$.  

Although $\mathbf{s}_q$ is unknown at the PUs, at the beginning of each IRS transmission frame, the AP emits some pilot symbols that allows the PUs to obtain the effective CSI at the receiver (CSIR) of the cascaded AP-to-IRS-to-PU$_k$ link with existing techniques, e.g. \cite{liu2020deep}.  As such, the PUs are able to apply coherent detection to decode their information symbols.
On the other hand, in contrast to the existing works, e.g.\cite{hua2021novel,yan2019passive,zhang2021reconfigurable,lin2020reconfigurable}, the SU is assumed to be a practical receiver which does not equip with powerful computational capability to perform joint decoding of $c_k(l)$ and ${\mathbf{s}_q}$. Without knowing AP symbols $c_k(l)$, the phase of the instantaneous CSI of the cascaded AP-to-IRS-to-SU link cannot be obtained for performing conventional coherent detection. In this case, a non-coherent detection \cite{tse2005fundamentals} is adopted at the SU for decoding ${\mathbf{s}_q}$.

\vspace*{-3mm}
\subsection{Signal Model}\vspace*{-1mm}
\subsubsection{Transmitted Signal at the AP}
For each AP transmission frame, the transmitted signal at the AP is given by
\begin{align}
\mathbf{c}(l)=\sum_{k\in\mathcal{K}} \mathbf{w}_kc_k(l), \forall l\in \mathcal{L}.\\[-8mm]\notag
\end{align}
\subsubsection{Reflected Signal at the IRS}\label{section:reflected_signal_IRS}
As only non-coherent detection can be performed at the SU, we introduce a multi-level amplitude modulation. In particular, the IRS consists of $Q>1$ information-carrying elements (ICEs) and $M-Q$ REs. With the similar idea as in \cite{5453306}, switching on/off among the ICEs can modulate the information into the signal swing of reflected signals from the IRS, by creating the required multiple power level signals. The remaining REs is in the ``on'' state at each IRS symbol duration to reflect the impinging signals. Note that ICEs in the ``on'' state reflect impinging signals as well but with its index modulated.
Based on this, the transmitted IRS symbol is denoted by $\mathbf{s}_q=[s_{1,q},...,s_{m,q},...,s_{M,q}]^{\mathrm{T}},\forall q\in\mathcal{Q}=\{1,\ldots,Q+1\}$. $s_{m,q}\in\{0,1\}$  denotes the on/off state of the $m$-th IRS element when IRS transmitting symbol $\mathbf{s}_q$, i.e., $s_{m,q}=0$ and $s_{m,q}=1$ denote that the $m$-th IRS elements is turned ``off'' and ``on'', respectively.  
 The truth table in Table \ref{tab:truth_table} shows an example of the proposed modulation scheme.
\begin{table}[t]
\small
\caption{\!A truth table of the proposed multilevel amplitude modulation scheme at the IRS with $Q\!=\!3$ ICEs and $M-Q$ REs.} \vspace{-3mm}
\begin{center}
\begin{tabular}{l c c c c}
\toprule
              & ICE 1 & ICE 2 & ICE 3 & All REs\\[-1mm]\midrule
 Level 1 & 0     & 0     & 0   & X\\
 Level 2 & 0     & 0     & 1   & X \\
 Level 3 & 0     & 1     & 1   &  X\\
 Level 4 & 1     & 1     & 1&X \\[-1mm]\bottomrule
\multicolumn{5}{l}{\small X = Don't care}
\end{tabular}
\label{tab:truth_table}
\end{center}\vspace{-10mm}
\end{table}
In this case, there are $Q+1$ possible on/off patterns. Without loss of generality, we assume that $(Q+1)$ is a power of $2$.  Hence, each IRS symbol contains $\log_2(Q+1)$ IRS information bits.
 Since all IRS reflection patterns are assumed to be transmitted equiprobably. We have $\mathrm{Pr}\{\mathcal{H}_q\}=\frac{1}{Q+1}$, where $\mathcal{H}_q$ is the hypothesis of the IRS sending symbol $\mathbf{s}_q$.

\subsubsection{Received Signal at PUs}
For each AP transmission frame, the received signal at $\mathrm{PU}$$_k$ is given by\vspace{-2mm}
\begin{align}
y_{\mathrm{P},k}(l)\hspace{-1mm}  = \hspace{-1mm}  \mathbf{h}_{\mathrm{P},k}^{\mathrm{H}}\mathbf{S}_q\mathbf{\Phi}\mathbf{G}\sum_{k\in \mathcal{K} }\mathbf{w}_kc_k(l)+n_k(l), \forall k, l,\\[-7mm] \notag
\end{align}
where $\mathbf{S}_q=\mathrm{diag}(\mathbf{s}_q)$ and $n_k(l)\sim\mathcal{CN}(0,\sigma_k^2)$ denotes the background noise at $\mathrm{PU}$$_k$ with power $\sigma_k^2$.
Since $\mathbf{s}_q$ remains unchanged during the $L$ AP symbol durations, the achievable rate for $\mathrm{PU}$$_k$ to decode $c_k(l)$ is given by\vspace*{-2mm}
\begin{align}
R_{q,k}^\mathrm{PU}\hspace{-1mm} =\hspace{-1mm} \log_2\hspace{-1mm} \Bigg( 1\hspace{-1mm} +\hspace{-1mm} \frac{|\mathbf{h}_{\mathrm{P},k}^{\mathrm{H}}\mathbf{S}_q\mathbf{\Phi}\mathbf{G}\mathbf{w}_k|^2}{\sum_{\substack{j\neq k}}^{K}|\mathbf{h}_{\mathrm{P},k}^{\mathrm{H}}\mathbf{S}_q\mathbf{\Phi}\mathbf{G}\mathbf{w}_j|^2\!+\!\sigma_k^2}\Bigg),\forall k, q,\label{eq:sinrK}
\end{align}
where $\mathbf{H}_{\mathrm{P},k}=\mathrm{diag}(\mathbf{h}_{\mathrm{P},k}^{\mathrm{H}})$.
 Although the CSIR of the cascaded AP-IRS-PU$_k$ link, i.e., $\mathbf{h}_{\mathrm{P},k}^{\mathrm{H}}\mathbf{S}_q\mathbf{\Phi}\mathbf{G}$, is known at the PU$_k$, the AP has no prior knowledge of $\mathbf{s}_q$ while optimizing $\mathbf{w}_k$ and $\bm{\Phi}$. Hence, to facilitate the precoding and IRS reflection coefficients design, the average achievable rate of $\mathrm{PU}$$_k$ is adopted. To this end, the expectation of $R_{q,k}^\mathrm{PU}$ is taken over $\mathbf{s}_q$ as \vspace*{-3mm}
\begin{align}
\overline{R}_{k}^\mathrm{PU}=\mathbb{E}_{\mathbf{s}_q}\Big[R_{q,k}^\mathrm{PU}\Big], \forall k.\label{eq:sinrK_E}\\[-7mm]\notag
\end{align}
\subsubsection{Received Signal at SU}
Since each IRS transmission frame spans $L$ AP symbol durations, the received signal at the SU for each IRS transmission frame is given by\vspace*{-2mm}
\begin{align}
\mathbf{y}_{\mathrm{S}} \hspace{-1mm}=&  [y_{\mathrm{S}}(1),\ldots,y_{\mathrm{S}}(l),\ldots,y_{\mathrm{S}}(L)]^{\mathrm{T}}\text{, with}\\
\hspace{-1mm}y_{\mathrm{S}}(l)\hspace{-1mm}=&  \mathbf{h}_{\mathrm{S}}^{\mathrm{H}}\mathbf{S}_q\mathbf{\Phi}\mathbf{G}\mathbf{c}(l)\hspace{-1mm}+\hspace{-1mm}n_{\mathrm{S}}(l)=
\mathbf{s}_q^{\mathrm{T}}\mathbf{\Phi}\mathbf{H}_{\mathrm{S}}\mathbf{G}\mathbf{c}(l)+n_{\mathrm{S}}(l),\forall l,\hspace{-1mm}\notag\\[-7mm]\notag
\end{align}
being the received signal of the SU at the $l$-th observation, where $\mathbf{H}_{\mathrm{S}}=\mathrm{diag}(\mathbf{h}_{\mathrm{S}}^{\mathrm{H}})$ and $n_{\mathrm{S}}(l)\sim\mathcal{CN}(0,\sigma_{\mathrm{S}}^2)$ denotes the noise at the SU with power $\sigma_{\mathrm{S}}^2$.

As the SU does not have the prior knowledge of $c_k(l)$, the effective CSIR of the cascaded AP-IRS-SU link, $\mathbf{\Phi}\mathbf{H}_{\mathrm{S}}\mathbf{G}\mathbf{c}(l)$, is not available at the SU.
 Fortunately, the distribution of the AP symbols is  $c_k(l)\sim\mathcal{CN}(0,1)$ such that the distribution of $y_{\mathrm{S}}(l)$ is available at the SU, i.e., $y_{\mathrm{S}}(l)\sim \mathcal{CN}(0,P_{\mathbf{s}_q}+\sigma^2_{\mathrm{s}})$, where $P_{\mathbf{s}_q}=\sum_{k=1}^K\mathrm{Tr}\Big(\mathbf{s}_q^{\mathrm{T}}
\mathbf{\Phi}\mathbf{H}_{\mathrm{S}}\mathbf{G}\mathbf{w}_k
\mathbf{w}_k^{\mathrm{H}}\mathbf{G}^{\mathrm{H}}\mathbf{H}_{\mathrm{S}}^{\mathrm{H}}\mathbf{\Phi}^{\mathrm{H}}\mathbf{s}_q\Big)$. As a result, the distribution of the received signal power at the SU is  $\sum_{l=1}^L y_{\mathrm{S}}^2(l)\sim \mathrm{Erlang}(\frac{1}{\lambda}_q,L)$ with $\lambda_q=( P_{\mathbf{s}_q}+\sigma^2_{\mathrm{s}})^{-1}, \forall q\in\mathcal{Q}
$.
Therefore, we adopt the non-coherent detection method\cite{tse2005fundamentals} to detect the modulated IRS symbols at the SU via\vspace*{-2mm}
\begin{align}
&[i]=\underset{q\in\mathcal{Q}  }{\arg \max}\,\,p\Big(\sum_{l\in\mathcal{L}} y_{\mathrm{S}}^2(l)|\mathcal{H}_q\Big)\mathrm{Pr}\{\mathcal{H}_q\},\text{ where}\label{eq:detector}\\
&p\Big(\sum_{l\in\mathcal{L}} y_{\mathrm{S}}^2(l)|\mathcal{H}_q\Big)\!\!=\!\!\frac{\lambda_{q}^{L}\big(\sum_{l\in \mathcal{L}}y_{\mathrm{S}}^2(l)\big)^{L-1}e^{-\lambda_{q}\sum_{l\in \mathcal{L}}y_{\mathrm{S}}^2(l)}}{(L-1)!}.\label{eq:pdf_ys}\\[-7mm]\notag
\end{align}
It can be observed from \eqref{eq:detector} that the non-coherent detection performance can be improved by exploiting $\lambda_q$ and $y_{\mathrm{S}}$ in \eqref{eq:pdf_ys}, which can be achieved by optimizing the precoder at the AP and IRS phase shifts.

To characterize the detection performance of the SU, we first derive an SER upper bound $P_{\mathrm{e}}^{\mathrm{Upper}}$, by adopting the well-known union bounding technique, whose tightness has been verified in \cite{jeganathan2008spatial}, that yields\vspace*{-2mm}
\begin{align}
&P_{\mathrm{e}}^{\mathrm{Upper}}=\sum_{\substack{q=1}}^{Q+1}\sum_{\substack{i=1,i\neq q}}^{Q+1}\mathrm{Pr}\Big\{\mathbf{s}_q\rightarrow\mathbf{s}_i|\mathcal{H}_q\Big\}
\mathrm{Pr}\{\mathcal{H}_q\},\label{Eq:errorbit_upper}\\[-7mm]\notag
\end{align}
where $\mathrm{Pr}\big\{\mathbf{s}_q\rightarrow\mathbf{s}_i|\mathcal{H}_q\big\}$ is the pairwise error probability  in deciding $\mathbf{s}_{q}$ to $\mathbf{s}_{i},\forall i\!\neq\! q,$ under the hypothesis $\mathcal{H}_q$. According \eqref{eq:detector}, the total error probability of the hypothesis $\mathcal{H}_q$ is \vspace*{-2mm}
\begin{align}
&\sum_{\substack{i=1,i\neq q}}^{Q+1}\mathrm{Pr}\Big\{\mathbf{s}_q\rightarrow\mathbf{s}_i|\mathcal{H}_q\Big\}\\[-1mm]
=&\begin{cases}
      \frac{1}{(L-1)!} \Gamma(L,L\lambda_q d_q), &q=1,\\[-1mm]
      \frac{1}{(L-1)!}\Big(\gamma(L,L\lambda_q d_{q-1})+\Gamma(L,L\lambda_q d_q)\Big), &1<q\leq Q,  \\[-1mm]
      \frac{1}{(L-1)!} \gamma(L,L\lambda_q d_{q-1}),  &q=Q+1.
    \end{cases}\notag\\[-7mm]\notag
\end{align}
Here, $\gamma(s,x)=\int^x_0 t^{s-1}e^{-t}dt$ and $\Gamma(s,x)=\int^{\infty}_x t^{s-1}e^{-t}dt$ are the lower incomplete Gamma function and the upper incomplete Gamma function, respectively. $d_q=\frac{1}{2}(P_{\mathbf{s}_{q+1}}-P_{\mathbf{s}_q})$ is the detection threshold between $\mathcal{H}_q$ and $\mathcal{H}_{q+1}$.
Therefore, we have\vspace{-2mm}
\begin{align}
&P_{\mathrm{e}}^{\mathrm{Upper}}\!\!=\!\! \frac{\sum_{\substack{\hat{q}=2}}^{Q+1}\gamma(L,L\lambda_{\hat{q}} d_{\hat{q}-1})+ \sum_{\substack{\tilde{q}=1}}^{Q}\Gamma(L,L\lambda_{\tilde{q}} d_{\tilde{q}})}{(Q+1)(L-1)!}.  \\[-7mm]\notag 
\end{align}
Since $L$ is an integer, we have $\gamma(L,L\lambda_{\hat{q}} d_{\hat{q}-1})=(L-1)!\Big(1-e^{-L\lambda_{\hat{q}} d_{{\hat{q}}-1}}\sum_{l=0}^{L-1}\frac{(L\lambda_{\hat{q}} d_{{\hat{q}}-1})^l}{l!}\Big)$ and $\Gamma(L,L\lambda_{\tilde{q}} d_{\tilde{q}})=(L-1)!e^{-L\lambda_{\tilde{q}} d_{{\tilde{q}}}}\sum_{l=0}^{L-1}\frac{(L\lambda_{\tilde{q}} d_{{\tilde{q}}})^l}{l!},\forall \hat{q}=\{2,\ldots,Q+1\},\tilde{q}=\{1,\ldots,Q\}$, respectively\cite[Th.~3, Th.~4]{jameson2016incomplete}.

\vspace*{-2mm}
\section{Problem Formulation}
We aim to maximize the average achievable sum-rate of the primary system while guaranteeing the power budget at the AP and the QoS requirements of the SU by jointly designing the precoding vectors $\mathbf{w}_{k},\forall k, $ and the phase shifts $\theta_m, \forall m$. The joint design can be formulated as the following:\vspace*{-2mm}
\begin{align}
\underset{\mathbf{w}_k,\, \theta_m\,}{\mathrm{maximize}} \,\,& \sum_{k\in\mathcal{K}}\overline{R}_{k}^\mathrm{PU}\label{eq:proposed_formulation_origion} \\[-2mm]
\mathrm{s.t.}\,\,&\mathrm{C1}\hspace{-1mm}:\sum_{k\in\mathcal{K}}\|\mathbf{w}_k\|^2\leq P_{\max}, \forall k\in\mathcal{K} ,\notag\\[-1mm]
&\mathrm{C2}\hspace{-1mm}: P_{\mathrm{e}}^{\mathrm{Upper}}\leq P_{\mathrm{e}}^{\max},
\mathrm{C3}\hspace{-1mm}:0 < \theta_m\leq 2\pi ,\forall m\in\mathcal{M}.\notag\\[-8mm]\notag
\end{align}
Constraint C1 ensures that the transmit power consumption of the precoder at the AP is less than the maximum available power budget $P_{\max}$. Constraint C2 is imposed to restrict the upper bound of the SER for decoding the modulated IRS symbols at the SU to be less than a constant $P_{\mathrm{e}}^{\max}$ defined by the target application. Constraint C3 specifies that $\theta_m$ can only vary from $0$ to $2\pi$. The formulated problem is non-convex due to the non-convexities in both the upper and lower incomplete Gamma
functions in constraint C2 and the couplings among optimization variables $\mathbf{w}_k$ and $\theta_m$ in both the objective function and constraint C2.
In general, the application of a brute-force search is required for obtaining the globally optimal solution of \eqref{eq:proposed_formulation_origion}, which is computationally prohibited even for a moderate system size. As an alternative, a computationally efficient suboptimal algorithm is proposed in the next section.
\vspace*{-2mm}
\section{Solution of the Optimization Problem}
To address the proposed optimization problem in \eqref{eq:proposed_formulation_origion}, we first transform the objective function and constraint C2 into their equivalent forms, such that they are convex with respect to (w.r.t.) $\mathbf{\Phi}\mathbf{H}_{\mathrm{P},k}\mathbf{G}\mathbf{w}_j,\forall k,j,$ and $\mathbf{\Phi}\mathbf{H}_{\mathrm{S}}\mathbf{G}\mathbf{w}_k,\forall k$, respectively. Then, we decouple the coupling between optimization variables $\mathbf{w}_k$ and $\theta_m$ by utilizing the Schur complement. Finally, SCA is applied to address the non-convex constraints in the transformed optimization problem.

As shown in \eqref{Eq:errorbit_upper} and  \eqref{eq:proposed_formulation_origion}, constraint C2 in \eqref{eq:proposed_formulation_origion} is non-convex due to the highly coupled variables.  We first introducing auxiliary optimization variables $\gamma_{\hat{q}}$, $\Gamma_{\tilde{q}}, \xi_{\hat{q}}^{\gamma},  \epsilon_{\hat{q}}^{\gamma 1},\epsilon_{\hat{q}}^{\gamma 2}$  $\xi_{\tilde{q}}^{\Gamma},\epsilon_{\tilde{q}}^{\Gamma 1}$, and $\epsilon_{\tilde{q}}^{\Gamma 2}, \forall \hat{q},\tilde{q},$ for decoupling. Then, by exploiting the properties of logarithm and quadratic equation, constraint C2  can be equivalently transformed as:\hspace{-4mm}
\begin{align}
\mathrm{C2a}\hspace{-1mm}:&\sum_{\hat{q}=2}^{Q+1}\gamma_{\hat{q}}+\sum_{\tilde{q}=1}^{Q}\Gamma_{\tilde{q}}\leq (Q+1)P_{\mathrm{e}}^{\max},\label{eq:C2a_c}\\[-2mm]
\mathrm{C2b1}\hspace{-1mm}:& \!-\!L\epsilon_{\hat{q}}^{\gamma 1}\!\!+\!\ln\xi_{\hat{q}}^{\gamma}\!\geq \!\ln(1\!-\!\gamma_{\hat{q}}),
\mathrm{C2b2}\hspace{-1mm}:\xi_{\hat{q}}^{\gamma}\!\leq\! \!\sum_{l=0}^{L\!-\!1}\!\frac{(L\epsilon_{\hat{q}}^{\gamma 2})^l}{l!},\!\forall {\hat{q}},\notag\\[-1mm]
\mathrm{C2b3}\hspace{-1mm}:&
P_{\mathbf{s}_{\hat{q}}}\!\!-\!\!P_{\mathbf{s}_{{\hat{q}}-1}}\!\!+\!\!\Big(\!\epsilon_{\hat{q}}^{\gamma 1}\!\!-\!\!P_{\mathbf{s}_{\hat{q}}}\!\!-\!\!\sigma^2_{\mathrm{s}}\Big)^2\!\!\!-\!\!(\epsilon_{\hat{q}}^{\gamma 1})^2\!\!-\!\!(P_{\mathbf{s}_{\hat{q}}}\!\!+\!\sigma^2_{\mathrm{s}}\!)^2\!\!\leq\!\!0,\forall {\hat{q}},\notag\\[-1mm]
\mathrm{C2b4}\hspace{-1mm}:
&
P_{\mathbf{s}_{\hat{q}}}\!\!-\!\!P_{\mathbf{s}_{{\hat{q}}-1}}
\!\!\!-\!\!\Big(\epsilon_{\hat{q}}^{\gamma 2}\!\!+\!\!P_{\mathbf{s}_{\hat{q}}}\!\!+\!\sigma^2_{\mathrm{s}}\Big)^2\!\!\!+\!(\epsilon_{\hat{q}}^{\gamma 2})^2\!\!\!+\!\!(P_{\mathbf{s}_{\hat{q}}}\!\!\!+\!\sigma^2_{\mathrm{s}})^2\!\!\geq\!\!0,\forall {\hat{q}},\notag\\[-1mm]
\mathrm{C2c1}\hspace{-1mm}:&-\!L\epsilon_{\tilde{q}}^{\Gamma 1} \!+ \!\ln\xi_{\tilde{q}}^{\Gamma}\leq\ln\Gamma_{\tilde{q}},
\mathrm{C2c2}\hspace{-1mm}:\sum_{l=0}^{L-1}\frac{(L\epsilon_{\tilde{q}}^{\Gamma 2})^l}{l!}\leq \xi_{\tilde{q}}^{\Gamma},\forall {\tilde{q}},\notag\\[-2mm]
\mathrm{C2c3}\hspace{-1mm}:& P_{\mathbf{s}_{{\tilde{q}}+1}}\!\!-\!\!P_{\mathbf{s}_{{\tilde{q}}}}\!\!\!-\!\!\Big(\!\epsilon_{\tilde{q}}^{\Gamma 1}\!\!+\!\!P_{\mathbf{s}_{\tilde{q}}}\!+\!\sigma^2_{\mathrm{s}}\Big)^2\!\!\!+\!\!(\epsilon_{\tilde{q}}^{\Gamma 1})^2\!\!\!+\!\!(P_{\mathbf{s}_{\tilde{q}}}\!\!+\!\!\sigma^2_{\mathrm{s}})^2\!\!\geq\!\!0,\forall {\tilde{q}}\notag\\[-1mm]
\mathrm{C2c4}\hspace{-1mm}:&
P_{\mathbf{s}_{{\tilde{q}}+1}}\!\!-\!\!P_{\mathbf{s}_{{\tilde{q}}}}\!\!+\!\!\Big(\epsilon_{\tilde{q}}^{\Gamma 2}\!\!\!-\!\!P_{\mathbf{s}_{\tilde{q}}}\!-\!\sigma^2_{\mathrm{s}}\Big)^2\!\!\!-\!\!(\epsilon_{\tilde{q}}^{\Gamma 2})^2\!\!\!-\!\!(P_{\mathbf{s}_{\tilde{q}}}\!\!+\!\!\sigma^2_{\mathrm{s}})^2\!\!\!\leq\!\!0,\forall {\tilde{q}}.\notag\\[-6mm]\notag
\end{align}
In particular,
 $\gamma_{\hat{q}}$, $\Gamma_{\tilde{q}}, \xi_{\hat{q}}^{\gamma}$, and  $\xi_{\tilde{q}}^{\Gamma}$ replace $\gamma(L,L\lambda_{\hat{q}} d_{\hat{q}-1})$, $\Gamma(L,L\lambda_{\tilde{q}} d_{\tilde{q}})$, $\sum_{l=0}^{L-1}\frac{(L\lambda_{\hat{q}} d_{{\hat{q}}-1})^l}{l!}$, and $\sum_{l=0}^{L-1}\frac{(L\lambda_{\tilde{q}} d_{{\tilde{q}}})^l}{l!}$ in constraint C2, respectively.
 Besides, $\epsilon_{\hat{q}}^{\gamma 1}$ and $\epsilon_{\tilde{q}}^{\Gamma 1}$ replace $\lambda_{\hat{q}} d_{{\hat{q}}-1}$ and  $\lambda_{\tilde{q}} d_{{\tilde{q}}}$ in $e^{-L\lambda_{\hat{q}} d_{{\hat{q}}-1}}$ and $e^{-L\lambda_{\tilde{q}} d_{{\tilde{q}}}}$ of constraint C2, respectively.
Also, $\epsilon_{\hat{q}}^{\gamma 2}$ and $\epsilon_{\tilde{q}}^{\Gamma 2}$  replace $\lambda_{\hat{q}} d_{{\hat{q}}-1}$ and  $\lambda_{\tilde{q}} d_{{\tilde{q}}}$ in $\sum_{l=0}^{L-1}\frac{(L\lambda_{\hat{q}} d_{{\hat{q}}-1})^l}{l!}$ and $\sum_{l=0}^{L-1}\frac{(L\lambda_{\tilde{q}} d_{{\tilde{q}}})^l}{l!}$ of constraint C2, respectively.
It can be verified that constraints C2b3, C2b4, C2c3, and C2c4 are convex w.r.t. $\mathbf{\Phi}\mathbf{H}_{\mathrm{S}}\mathbf{G}\mathbf{w}_k$ that paves the way for further simplification in the sequel.

On the other hand, since $\mathbf{s}_q$ has a finite number of choices, i.e., $Q+1$, the average achievable capacity in \eqref{eq:sinrK_E} can be directly expressed as
$
\overline{R}_{k}^\mathrm{PU}=\sum_{q=1}^{Q+1}R_{q,k}^\mathrm{PU}\mathrm{Pr}\{\mathcal{H}_q\}
$. Thus, the objective function in \eqref{eq:proposed_formulation_origion} can be equivalently transformed to a new objective function as\vspace{-2mm}
\begin{align}
\hspace{-1mm}f_{\mathrm{o}}\!\!=\!\!\frac{1}{Q\!\!+\!\!1}\!\!\!\sum_{q=1}^{Q+1}\!\!\sum_{k=1}^K\!\!\Bigg(\!\!\!\log_2\hspace{-1mm}\Big(\hspace{-1mm}\sum_{j=1}^K\hspace{-1mm}P_{q,k,j}^{\mathrm{PU}}\hspace{-1mm}+\hspace{-1mm}\sigma_{k}^2\!\Big)\hspace{-1mm}\!-\hspace{-1mm}\log_2\!\!\Big(\hspace{-1mm}\sum_{j\neq1}^K\hspace{-1mm}P_{q,k,j}^{\mathrm{PU}}\hspace{-1mm}
+\hspace{-1mm}\sigma_{k}^2\!\Big)\!\!\!\Bigg)\!,\!\!\!\label{eq:New_OF}\\[-6mm]\notag
\end{align}%
where $P_{q,k,j}^{\mathrm{PU}}=\mathrm{Tr}(\mathbf{s}_q^{\mathrm{T}} \mathbf{\Phi}\mathbf{H}_{\mathrm{P},k}\mathbf{G}\mathbf{w}_j
\mathbf{w}_j^{\mathrm{H}}\mathbf{G}^{\mathrm{H}}\mathbf{H}_{\mathrm{P},k}^{\mathrm{H}}\mathbf{\Phi}^{\mathrm{H}}\mathbf{s}_q),\forall k,j,q$.

Next, we decouple the coupled variables $\{\mathbf{w}_k,\bm{\Phi}\}$ in the objective function in \eqref{eq:New_OF} via utilizing the Schur complement.
We first introduce auxiliary optimization variables $\mathbf{f}_{k,j}^{\mathrm{PU}}\! \in \! \mathbb{C}^{M\! \times\! 1}$, $\! \mathbf{U}_{k,j}^{\mathrm{PU}}\! \in \! \mathbb{C}^{M \! \times\!  M}\! $, $\! {c}_{k,j}^{\mathrm{PU}}, \forall k,j$, and $\mathbf{A}\! \in\!  \mathbb{C}^{M\! \times\!  M}$. Then, by substituting $\mathbf{\Phi}\mathbf{H}_{\mathrm{P},k}\mathbf{G}\mathbf{w}_j
\mathbf{w}_j^{\mathrm{H}}\mathbf{G}^{\mathrm{H}}\mathbf{H}_{\mathrm{P},k}^{\mathrm{H}}\mathbf{\Phi}^{\mathrm{H}}=\mathbf{U}_{k,j}^{\mathrm{PU}}$ into \eqref{eq:New_OF}, the objective function in \eqref{eq:New_OF} and constraint C3 in \eqref{eq:proposed_formulation_origion} can be equivalently transformed as \vspace{-2mm}
\begin{align}
&\overline{f_{\mathrm{o}}} = f_{\mathrm{o}}\Big|_{\mathbf{\Phi}\mathbf{H}_{\mathrm{P},k}\mathbf{G}\mathbf{w}_j
\mathbf{w}_j^{\mathrm{H}}\mathbf{G}^{\mathrm{H}}\mathbf{H}_{\mathrm{P},k}^{\mathrm{H}}\mathbf{\Phi}^{\mathrm{H}}=\mathbf{U}_{k,j}^{\mathrm{PU}}},\label{eq:New_OF_1}\\
&\mathrm{C4a}\hspace{-1mm}:\begin{bmatrix}\mathbf{U}_{k,j}^{\mathrm{PU}}&\mathbf{f}_{k,j}^{\mathrm{PU}}\\(\mathbf{f}_{k,j}^{\mathrm{PU}})^{\mathrm{H}}&1\end{bmatrix}\succeq \mathbf{0},\forall k,j,\notag\\[-1mm]
&\mathrm{C4b}\hspace{-1mm}:\mathrm{Tr}(\mathbf{U}_{k,j}^{\mathrm{PU}})-\mathrm{Tr}\Big(\mathbf{f}_{k,j}^{\mathrm{PU}}(\mathbf{f}_{k,j}^{\mathrm{PU}})^{\mathrm{H}}\Big)\leq0,\forall k,j,\notag\\[-1mm]
&\mathrm{C4c}\hspace{-1mm}:\begin{bmatrix}\mathbf{D}_{k,j}^{\mathrm{PU}}&\mathbf{E}_{k,j}^{\mathrm{PU}}\\[0.5mm](\mathbf{E}_{k,j}^{\mathrm{PU}})^{\mathrm{H}}&\mathbf{I}_M\end{bmatrix}\succeq \mathbf{0},\forall k,j,\notag\\[-1mm]
&\mathrm{C4d}\hspace{-1mm}:\mathrm{Tr}(\mathbf{D}_{k,j}^{\mathrm{PU}}) -\mathrm{Tr}\Big(\mathbf{E}_{k,j}^{\mathrm{PU}} (\mathbf{E}_{k,j}^{\mathrm{PU}})^{\mathrm{H}}\Big)\leq 0,\forall k,j,\notag\\[-1mm]
&\mathrm{C3a}\hspace{-1mm}: |\mathbf{\Phi}(m,m)|\leq1, \forall m \text{, and }\mathrm{C3b}\hspace{-1mm}:\mathbf{A}(m,m)\geq1, \forall m,\notag\\[-8mm]\notag
\end{align}
respectively, where $\mathbf{D}_{k,j}^{\mathrm{PU}}=\begin{bmatrix}\mathbf{A}& \mathbf{f}_{k,j}^{\mathrm{PU}}\\[0.5mm](\mathbf{f}_{k,j}^{\mathrm{PU}})^{\mathrm{H}}&{c}_{k,j}^{\mathrm{PU}}\end{bmatrix}$ and $\mathbf{E}_{k,j}^{\mathrm{PU}}=\begin{bmatrix}\bm{\Phi} \\(\mathbf{H}_{\mathrm{P},k}\mathbf{G}\mathbf{w}_k)^\mathrm{H}\end{bmatrix}$.

According to the Schur complement\cite[Th.~1.12]{zhang2006schur}, constraints C4a and C4b ensure that $\mathbf{U}_{k,j} = \mathbf{f}_{k,j}^{\mathrm{PU}}(\mathbf{f}_{k,j}^{\mathrm{PU}})^{\mathrm{H}}$ holds. Similarly, by combining constraints C3b, C4c, and C4d, we have  $\mathbf{f}_{k,j} = \mathbf{\Phi}\mathbf{H}_{\mathrm{P},k}\mathbf{G}\mathbf{w}_j$ and $\mathbf{A}=\mathbf{\Phi}\mathbf{\Phi}^{\mathrm{H}}$, such that  $\mathbf{U}_{k,j}^{\mathrm{PU}}$ is equivalent to $\mathbf{\Phi}\mathbf{H}_{\mathrm{P},k}\mathbf{G}\mathbf{w}_j\mathbf{w}_j^{\mathrm{H}}\mathbf{G}^{\mathrm{H}}\mathbf{H}_{\mathrm{P},k}^{\mathrm{H}}\mathbf{\Phi}^{\mathrm{H}}$ in \eqref{eq:New_OF}.

Likewise, by introducing auxiliary optimization variables $\mathbf{f}_{k}^{\mathrm{SU}}\in\mathbb{C}^{M\times 1}$, $\mathbf{U}_{k}^{\mathrm{SU}}\in\mathbb{C}^{M\times M}$, and ${c}_{k}^{\mathrm{SU}}$, constraints C2b3, C2b4, C2c3, and C2c4 can be equivalently transformed as \eqref{eq:SU_schur} via substituting ${P}_{\mathbf{s}_p}=\widetilde{P}_{\mathbf{s}_p}$ into \eqref{eq:C2a_c}:\vspace*{-1mm}
\begin{align}
\overline{\mathrm{C2b3}}\hspace{-1mm}:& \mathrm{C2b3}\Big|\!_{{P}_{\mathbf{s}_{\hat{q}}}\!=\!\widetilde{P}_{\mathbf{s}_{\hat{q}}},{P}_{\mathbf{s}_{\hat{q}\!-\!1}}\!=\!\widetilde{P}_{\mathbf{s}_{\hat{q}\!-\!1}}}\!\!,
\overline{\mathrm{C2b4}}\hspace{-1mm}: \! \mathrm{C2b4}\Big|\!_{{P}_{\mathbf{s}_{\hat{q}}}\!=\!\widetilde{P}_{\mathbf{s}_{\hat{q}}},{P}_{\mathbf{s}_{\hat{q}\!-\!1}}\!=\!\widetilde{P}_{\mathbf{s}_{\hat{q}\!-\!1}}},\notag\\
\overline{\mathrm{C2c3}}\hspace{-1mm}:& \mathrm{C2b3}\Big|\!_{{P}_{\mathbf{s}_{\tilde{q}}}\!=\!\widetilde{P}_{\mathbf{s}_{\tilde{q}}},{P}_{\mathbf{s}_{\tilde{q}\!+1\!}}\!=\!\widetilde{P}_{\mathbf{s}_{\tilde{q}\!+\!1}}},
\overline{\mathrm{C2c4}}\hspace{-1mm}:\!\mathrm{C2b4}\Big|\!_{{P}_{\mathbf{s}_{\tilde{q}}}\!=\!\widetilde{P}_{\mathbf{s}_{\tilde{q}}},{P}_{\mathbf{s}_{\tilde{q}\!+\!1}}\!=\!\widetilde{P}_{\mathbf{s}_{\tilde{q}\!+\!1}}},\notag\\[-0mm]
\mathrm{C2d1}\hspace{-1mm}:&\hspace{-1mm}\begin{bmatrix}\mathbf{U}_{k}^{\mathrm{SU}}&\hspace{-3mm}\mathbf{f}_{k}^{\mathrm{SU}}\\(\mathbf{f}_{k}^{\mathrm{SU}})^{\mathrm{H}}&\hspace{-3mm}1\end{bmatrix}\hspace{-1mm}\succeq \hspace{-1mm}\mathbf{0},
\mathrm{C2d2}\hspace{-1mm}:\hspace{-1mm}\mathrm{Tr}(\mathbf{U}_{k}^{\mathrm{SU}}\!)\hspace{-1mm}-
\hspace{-1mm}\mathrm{Tr}\Big(\hspace{-1mm}\mathbf{f}_{k}^{\mathrm{SU}}\!(\mathbf{f}_{k}^{\mathrm{SU}}\!)^{\mathrm{H}}\Big)\hspace{-1mm}\leq\hspace{-1mm}0,\forall k,\notag\\[-0mm]
\mathrm{C2d3}\hspace{-1mm}:&\begin{bmatrix}\mathbf{D}_{k}^{\mathrm{SU}}&\mathbf{E}_{k}^{\mathrm{SU}}\\(\mathbf{E}_{k}^{\mathrm{SU}})^{\mathrm{H}}&\mathbf{I}_{M}\end{bmatrix}\succeq \mathbf{0}, \forall k \text{, and }\notag\\[-1mm]
\mathrm{C2d4}\hspace{-1mm}:&\mathrm{Tr}(\mathbf{D}_{k}^{\mathrm{SU}}) -\mathrm{Tr}\Big(\mathbf{E}_{k}^{\mathrm{SU}} (\mathbf{E}_{k}^{\mathrm{SU}})^{\mathrm{H}}\Big)\leq 0,\forall k,\label{eq:SU_schur}\\[-7mm]\notag
%
\end{align}
where $\mathbf{D}_{k}^{\mathrm{SU}}=\begin{bmatrix}\mathbf{A}& \mathbf{f}_{k}^{\mathrm{SU}}\\(\mathbf{f}_{k}^{\mathrm{SU}})^{\mathrm{H}}&{c}_{k}^{\mathrm{SU}}\end{bmatrix}$, $\mathbf{E}_{k}^{\mathrm{SU}}=\begin{bmatrix}\bm{\Phi} \\ (\mathbf{H}_{\mathrm{S}}\mathbf{G}\mathbf{w}_k)^\mathrm{H}\end{bmatrix}$,\vspace{1mm} and $\widetilde{P}_{\mathbf{s}_p}=\sum_{k=1}^K\mathrm{Tr}\Big(\mathbf{s}_p^{\mathrm{T}}
\mathbf{U}^{\mathrm{SU}}_{k}\mathbf{s}_p\Big), p=\{\hat{q},\tilde{q}\}$.

\begin{figure*}
\small
\begin{align}
 &\underset{
\substack{\mathbf{w}_k,\, \mathbf{\theta}_m\,,\mathcal{A}}
}{\mathrm{maximize}} \,\,\frac{1}{Q\!\!+\!\!1}\sum_{q=1}^{Q+1}\!\sum_{k=1}^K\!\Bigg(\!\log_2\hspace{-1mm}\Big(\hspace{-1mm}\sum_{j=1}^K\hspace{-1mm}P_{q,k,j}^{\mathrm{PU}}\hspace{-1mm}+\hspace{-1mm}\sigma_{k}^2\Big)\hspace{-1mm}-\log_2\Big(\sum_{j\neq1}^KP_{q,k,j}^{\mathrm{PU}(\tau)}+\sigma_{k}^2\Big)
-\sum_{j\neq k}^K(P_{q,k,j}^{\mathrm{PU}}-P_{q,k,j}^{\mathrm{PU}(\tau)})/\Big(\ln(2)(\sum_{i\neq j}^KP_{q,j,i}^{\mathrm{PU}(\tau)}+\sigma_{k}^2)\Big)\hspace{-1mm}
\Bigg)\notag\\[-1mm]
&\mathrm{s.t.}\,\, \mathrm{C1},\mathrm{C2a},\overline{\mathrm{C2b1}},\mathrm{C2c2}, \mathrm{C2d1},
\mathrm{C2d3},\mathrm{C3a},\mathrm{C3b},\mathrm{C4a},\mathrm{C4c},\label{eq:proposed_formulation_sca}\tag{19} \\[-1mm]
&\overline{\mathrm{C2b2}}\hspace{-1mm}:\xi_{\hat{q}}^{\gamma}\!\leq \!\sum_{l=0}^{L-1}\frac{(L\epsilon_{\hat{q}}^{\gamma 2(\tau)})^l}{l!} +
\sum_{l=0}^{L-1}\frac{l(L\epsilon_{\hat{q}}^{\gamma 2(\tau)})^{(l-1)}L}{l!}\Big(\epsilon_{\hat{q}}^{\gamma 2}-\epsilon_{\hat{q}}^{\gamma 2(\tau)}\Big),\forall {\hat{q}},\notag\\[-2mm]
&\overline{ \overline{\mathrm{C2b3}}}\hspace{-1mm}:\widetilde{P}_{\mathbf{s}_{\hat{q}}}\!-\!\widetilde{P}_{\mathbf{s}_{{\hat{q}}-1}}\!+\!\Big(\epsilon_{\hat{q}}^{\gamma 1}\!-\!\widetilde{P}_{\mathbf{s}_{\hat{q}}}\!-\!\sigma^2_{\mathrm{s}}\Big)^2\!-\!(\epsilon_{\hat{q}}^{\gamma 1(\tau)})^2+2\epsilon_{\hat{q}}^{\gamma 1(\tau)}(\epsilon_{\hat{q}}^{\gamma 1}-\epsilon_{\hat{q}}^{\gamma 1(\tau)})-(\widetilde{P}_{\mathbf{s}_{\hat{q}}}^{(\tau)}+\sigma^2_{\mathrm{s}})^2-2(\widetilde{P}_{\mathbf{s}_{\hat{q}}}^{(\tau)}
+\sigma^2_{\mathrm{s}})(\widetilde{P}_{\mathbf{s}_{\hat{q}}}-\widetilde{P}_{\mathbf{s}_{\hat{q}}}^{(\tau)})\!\leq\!0,\forall {\hat{q}},\notag\\[-1mm]
&\overline{\overline{\mathrm{C2b4}}}\hspace{-1mm}:
\widetilde{P}_{\mathbf{s}_{\hat{q}}}\!-\!\widetilde{P}_{\mathbf{s}_{{\hat{q}}-1}}\!-\!\Big(\epsilon_{\hat{q}}^{\gamma 2}\!+\!\widetilde{P}_{\mathbf{s}_{\hat{q}}}\!+\!\sigma^2_{\mathrm{s}}\Big)^2\!+\!(\epsilon_{\hat{q}}^{\gamma 2(\tau)})^2+2\epsilon_{\hat{q}}^{\gamma 2(\tau)}(\epsilon_{\hat{q}}^{\gamma 2}-\epsilon_{\hat{q}}^{\gamma 2(\tau)})\!+\!(\widetilde{P}_{\mathbf{s}_{\hat{q}}}^{(\tau)}+\sigma^2_{\mathrm{s}})^2+2(\widetilde{P}_{\mathbf{s}_{\hat{q}}}^{(\tau)}
+\sigma^2_{\mathrm{s}})(\widetilde{P}_{\mathbf{s}_{\hat{q}}}-\widetilde{P}_{\mathbf{s}_{\hat{q}}}^{(\tau)})\!\geq\!0,\forall {\hat{q}},\notag\\[-1mm]
&\overline{\mathrm{C2c1}}\hspace{-1mm}:-\!L\epsilon_{\tilde{q}}^{\Gamma 1}\! +\! \ln(\xi_{\tilde{q}}^{\Gamma(\tau)})+(\xi_{\tilde{q}}^{\Gamma}-\xi_{\tilde{q}}^{\Gamma(\tau)})/\xi_{\tilde{q}}^{\Gamma(\tau)}\!-\!\ln\Gamma_{\tilde{q}}\!\leq\!0,\forall {\tilde{q}},\notag\\[-1mm]
&\overline{\overline{\mathrm{C2c3}}}\hspace{-1mm}:\widetilde{P}_{\mathbf{s}_{\tilde{q}+1}}\!-\!\widetilde{P}_{\mathbf{s}_{\tilde{q}}} \!-\!\Big(\epsilon_{\tilde{q}}^{\Gamma 1}\!+\!\widetilde{P}_{\mathbf{s}_{\tilde{q}}}\!+\!\sigma^2_{\mathrm{s}}\Big)^2+(\epsilon_{\tilde{q}}^{\Gamma1(\tau)})^2+2\epsilon_{\tilde{q}}^{\Gamma1(\tau)}(\epsilon_{\tilde{q}}^{\Gamma1}
-\epsilon_{\tilde{q}}^{\Gamma1(\tau)})
+(\widetilde{P}_{\mathbf{s}_{\tilde{q}}}^{(\tau)}+\sigma^2_{\mathrm{s}})^2+2(\widetilde{P}_{\mathbf{s}_{\tilde{q}}}^{(\tau)}
+\sigma^2_{\mathrm{s}})(\widetilde{P}_{\mathbf{s}_{\tilde{q}}}-\widetilde{P}_{\mathbf{s}_{\tilde{q}}}^{(\tau)})\!\geq\!0,\forall {\tilde{q}},\notag\\[-1mm]
%
&\overline{\overline{\mathrm{C2c4}}}\hspace{-1mm}:\widetilde{P}_{\mathbf{s}_{{\tilde{q}}+1}}\!-\!\widetilde{P}_{\mathbf{s}_{\tilde{q}}}\!+\!\Big(\epsilon_{\tilde{q}}^{\Gamma 2}\!-\!\widetilde{P}_{\mathbf{s}_{\tilde{q}}}\!-\!\sigma^2_{\mathrm{s}}\Big)^2\!-\!(\epsilon_{\tilde{q}}^{\Gamma 2(\tau)})^2-2\epsilon_{\tilde{q}}^{\Gamma 2(\tau)}(\epsilon_{\tilde{q}}^{\Gamma 2}
-\epsilon_{\tilde{q}}^{\Gamma 2(\tau)})\!-\!(\widetilde{P}_{\mathbf{s}_{\tilde{q}}}^{(\tau)}+\sigma^2_{\mathrm{s}})^2-2(\widetilde{P}_{\mathbf{s}_{\tilde{q}}}^{(\tau)}
+\sigma^2_{\mathrm{s}})(\widetilde{P}_{\mathbf{s}_{\tilde{q}}}-\widetilde{P}_{\mathbf{s}_{\tilde{q}}}^{(\tau)})\!\leq\!0,\forall {\tilde{q}},\notag\\[-1mm]
&\overline{\mathrm{C2d2}}\hspace{-1mm}:\mathrm{Tr}(\mathbf{U}_{k}^{\mathrm{SU}})\!\leq\!-\|\mathbf{f}^{\mathrm{SU}(\tau)}_{k}\|^2
+2\mathrm{Tr}\Big((\mathbf{f}^{\mathrm{SU}(\tau)}_{k})^{\mathrm{H}}\mathbf{f}^{\mathrm{PU}}_{k}\Big),
\overline{\mathrm{C2d4}}\hspace{-1mm}:\mathrm{Tr}(\mathbf{D}_{k}^{\mathrm{SU}})\!\leq\! -\|\mathbf{E}^{\mathrm{SU}(\tau)}_{k}\|^2_{\mathrm{F}}
\hspace{-1mm}+\hspace{-1mm}2\mathrm{Tr}\Big((\mathbf{E}^{\mathrm{SU}(\tau)}_{k})^{\mathrm{H}}\mathbf{E}^{\mathrm{PU}}_{k}\Big),\forall k,\notag\\[-1mm]
& \overline{\mathrm{C4b}}\hspace{-1mm}:\mathrm{Tr}(\mathbf{U}_{k,j}^{\mathrm{PU}}\!)\!\leq\!\hspace{-1mm}-\|\mathbf{f}^{\mathrm{PU}(\tau)}_{k,j}\|^2
+2\mathrm{Tr}\Big((\mathbf{f}^{\mathrm{PU}(\tau)}_{k,j})^{\mathrm{H}}\mathbf{f}^{\mathrm{PU}}_{k,j}\Big),
\overline{\mathrm{C4d}}\hspace{-1mm}:\!\mathrm{Tr}(\mathbf{D}_{k,j}^{\mathrm{PU}})\!\leq\! \hspace{-1mm}-\|\mathbf{E}^{\mathrm{PU}(\tau)}_{k,j}\|^2_{\mathrm{F}}
+2\mathrm{Tr}\Big((\mathbf{E}^{\mathrm{PU}(\tau)}_{k,j})^{\mathrm{H}}\mathbf{E}^{\mathrm{PU}}_{k,j}\Big),\forall k,j.\notag\\[-8mm]\notag
\end{align}
\hrulefill\vspace{-7mm}
\end{figure*}
For ease of presentation, we define a set $\mathcal{A}=\Big\{\mathbf{f}_{k}^{\mathrm{SU}}, \mathbf{U}_{k}^{\mathrm{SU}},  {c}_{k}^{\mathrm{SU}},
\mathbf{f}_{k,j}^{\mathrm{PU}},\mathbf{U}_{k,j}^{\mathrm{PU}}, {c}_{k,j}^{\mathrm{PU}},\mathbf{A}, \gamma_{\hat{q}},\Gamma_{\tilde{q}},\xi_{\hat{q}}^{\gamma}, \xi_{\tilde{q}}^{\Gamma},\epsilon_{\hat{q}}^{\gamma 1},\epsilon_{\tilde{q}}^{\Gamma 1}, \epsilon_{\hat{q}}^{\gamma 2},$ $\epsilon_{\tilde{q}}^{\Gamma 2}\Big\}$, which includes all introduced auxiliary optimization variables. Now, the optimization problem in \eqref{eq:proposed_formulation_origion} can be equivalently transformed to the following\vspace{-2mm}
\begin{align}
&\underset{
\substack{\mathbf{w}_k, \mathbf{\theta}_m,\mathcal{A}}
}{\mathrm{maximize}} \,\, \,\, \,\, \overline{f_{\mathrm{o}}} \notag\\[-1mm]
&\mathrm{s.t.}\,\,\,\, \mathrm{C1},\mathrm{C2a},\mathrm{C2b1},\mathrm{C2b2},\overline{\mathrm{C2b3}}, \overline{\mathrm{C2b4}},
 \mathrm{C2c1},\mathrm{C2c2}, \overline{\mathrm{C2c3}},\notag\\& \overline{\mathrm{C2c4}}, \mathrm{C2d1}-\mathrm{C2d4},
\mathrm{C3a},\mathrm{C3b},\mathrm{C4a}-\mathrm{C4d}.\label{eq:proposed_formulation_v1}\\[-8mm]\notag
\end{align}
Since constraints C2b1, C2b2, $\overline{\mathrm{C2b3}}$, $\overline{\mathrm{C2b4}}$, C2c1, $\overline{\mathrm{C2c3}}$, $\overline{\mathrm{C2c4}}$, C2d2, C2d4, C4b, C4d and the objective function in problem \eqref{eq:proposed_formulation_v1} are in the difference of convex (D.C.) functions form and differentiable, we apply an iterative method based on SCA to obtain a suboptimal solution. Taking constraint C2b1 as an example,  for any feasible point $\gamma_{\hat{q}}^{(\tau)}$, an upper bound of $\ln(1-\gamma_{\hat{q}})$ can be construct by deriving its first-order Taylor expansions:\vspace{-3mm}
\begin{align}
\ln(1-\gamma_{\hat{q}})\leq &
\Upsilon^1_{\hat{q}}\!=\!\ln(1-\gamma_{\hat{q}}^{(\tau)})\!+\!(\gamma_{\hat{q}}\!-\!\gamma_{\hat{q}}^{(\tau)})/(\gamma_{\hat{q}}^{(\tau)}\!-\!1),\\[-7mm]\notag
\end{align}
where $(\tau)$ denotes the iteration index for the proposed algorithm summarized in \textbf{Algorithm \ref{alg}} (to be discussed in detail later).
By applying SCA, a subset of constraint C2b1 can be obtained, which is given by \vspace*{-2mm}
    \begin{align}
\hspace{-3mm}&\overline{\mathrm{C2b1}}\hspace{-1mm}: -L\epsilon_{\hat{q}}^{\gamma 1}\!+\!\ln\xi_{\hat{q}}^{\gamma}\!- \!\Upsilon^1_{\hat{q}}\!\geq\!0,\forall {\hat{q}}.\\[-7mm]\notag
\end{align}
As $\overline{\mathrm{C2b1}}$ implies ${\mathrm{C2b1}}$,  replacing  ${\mathrm{C2b1}}$ with $\overline{\mathrm{C2b1}}$ can ensure
that the former is satisfied when the proposed algorithm converges.
Similarly, by applying SCA to the rest of D.C. functions in problem \eqref{eq:proposed_formulation_v1}, a lower bound of \eqref{eq:proposed_formulation_v1} can be obtained via solving the optimization problem in \eqref{eq:proposed_formulation_sca} at the top of this page.
\begin{table}[t] \vspace*{-2mm}
\scriptsize
\linespread{1.05}
\begin{algorithm} [H]
\caption{\small Proposed Suboptimal Resource Allocation Scheme } \label{alg}
\begin{algorithmic} [1]
\STATE Initialize the maximum number of iteration $(\tau)_{\max}$, the initial iteration index $\tau=0$, and optimization variables in $\mathcal{D}^{(\tau)}\!\!\! = \!\!\!\Big\{P_{q,k,j}^{\mathrm{PU}(\tau)}$,\!\!\! $\epsilon_{\hat{q}}^{\gamma 2(\tau)}$, $\widetilde{P}_{\mathbf{s}_p}^{(\tau)}$, $\epsilon_p^{i(\tau)}$, $\gamma_{\hat{q}}^{(\tau)}$, $\xi_{\tilde{q}}^{\Gamma(\tau)}$, $\mathbf{f}^{\mathrm{PU}(\tau)}_{k,j}, \mathbf{E}^{\mathrm{PU}(\tau)}_{k,j}, \mathbf{f}^{\mathrm{SU}(\tau)}_{k}, \mathbf{E}^{\mathrm{SU}(\tau)}_{k},\forall q,p,k,j,i,\hat{q},\tilde{q}\Big\}$.
\REPEAT[Main Loop: SCA]
\STATE Solve problem \eqref{eq:proposed_formulation_sca} with given optimization variables in $\mathcal{D}^{(\tau)}$, to obtain the variables for $\mathcal{D}^{(\tau+1)}$;
\STATE Set $\tau=\tau+1$ and update the optimization variables;
\UNTIL{convergence or $\tau=\tau_{\max}$}.\vspace*{-1mm}
\end{algorithmic}
\end{algorithm}
\vspace*{-9mm}
\end{table}
To tighten the obtained performance lower bound, we iteratively update the feasible solution by solving the optimization problem in \eqref{eq:proposed_formulation_sca} in the $(\tau)$-th iteration. The proposed SCA-based algorithm is shown in \textbf{Algorithm \ref{alg}} and the proof of its convergence to a suboptimal solution can be found in \cite{opial1967weak} which is omitted here for brevity. Note that the proposed algorithm has a polynomial time complexity.
\vspace{-3mm}
\section{Numerical Results}\vspace{-1mm}
This section evaluates the system performance of the proposed scheme via simulation. We set $K = 2$, $Q = 3$, $N_{\mathrm{t}} = 4$, $M = 30$, and $L = 30$. The location of AP, IRS, SU, and PUs are set in a Cartesian coordinate system, i.e., $(0,0)$, $(15,10)$, $(20,2)$, and $\{(65,2), (65,-2)\}$ in meters (m), respectively. The distance-dependent path loss model in \cite{wu2019intelligent} is adopted with a reference distance of $1$ m. Other important parameters are summarized as follows unless specified otherwise.  The centre carrier frequency is set as $2.4$ GHz. The path loss exponents of AP-IRS, IRS-SU, and IRS-PU$_k$ links are identical for simplicity, i.e., $\alpha_{\mathrm{AI}}=\alpha_{\mathrm{IS}}=\alpha_{\mathrm{IP},k}=2.2$. Rician factors of AP-IRS, IRS-SU, and IRS-PU$_k$ links are $\beta_{\mathrm{AI}}=\beta_{\mathrm{IS}}=\beta_{\mathrm{IP},k}=3$. The maximum power budget at the AP is $P_{\max} = 30$ dBm. Noise power at the SU and PUs are $\sigma^2_{k} = \sigma^2_{\mathrm{s}} = -100$ dBm.

For comparison, we also evaluate the system performance of three other schemes: 1) Baseline scheme 1 is identical to the proposed scheme except that the QoS of the SU constraints is not considered; 2) Baseline scheme 2 is the same as the proposed scheme except that the phase shifts of the reflect elements are randomly set; 3) A performance upper bound is achieved by an conventional IRS-assisted system with all elements being ``on'' state, while the SU does not exist. Note that except the upper bound scheme, for all the schemes, if the joint designed  precoder and IRS phase shifts are unable to meet QoS requirements of constraint C2 in \eqref{eq:proposed_formulation_origion}, we set the system sum-rate for that channel realization as zero to account the penalty for the corresponding failure.

\begin{figure}[t] \vspace*{-2mm}
  \centering
  \includegraphics[width=3.3in]{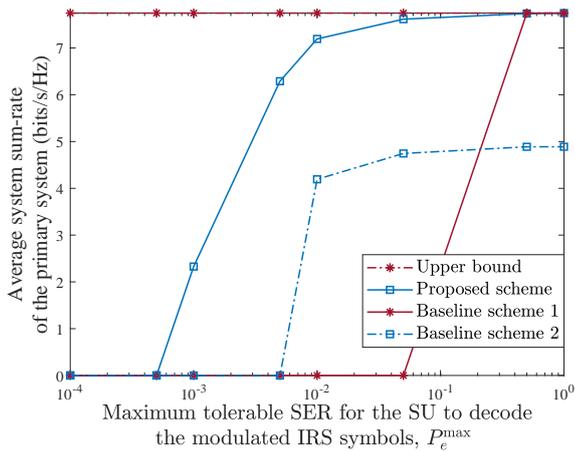}  \vspace*{-2mm}
  \caption{\hspace{-1mm} Average system achievable sum-rate of the primary system versus the upper bound of the SER at the SU side, $P^{\max}_{\mathrm{e}}$.}
  \label{Fig:SumRVSPeMax} \vspace*{-5mm}
\end{figure}

\begin{figure}[t] \vspace*{-2mm}
  \centering
  \includegraphics[width=3.3in]{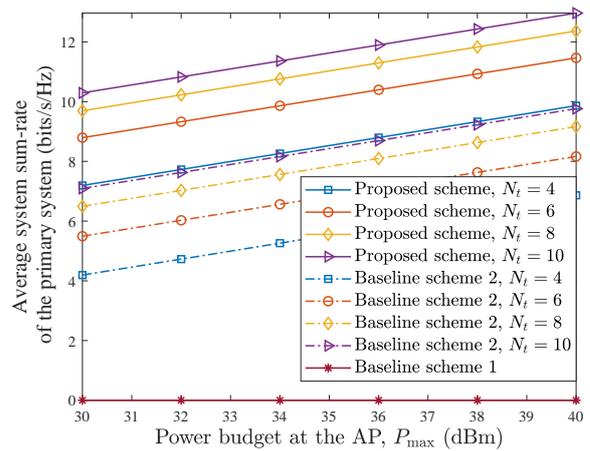}  \vspace*{-2mm}
  \caption{\hspace{-1mm} Average system achievable sum-rate of the primary system versus power budget, $P_{\max}$, with $P_{\mathrm{e}}^{\max}=0.01$.}
  \label{Fig:SumRVSP} \vspace*{-5mm}
\end{figure}

Fig. \ref{Fig:SumRVSPeMax} depicts the average system sum-rate of primary system versus the maximum tolerable SER, $P^{\max}_{\mathrm{e}}$, for the SU to decode the modulated IRS symbols. It can be observed that when $P^{\max}_{\mathrm{e}}$ is small, except the upper bound scheme, the average system sum-rates of all the considered schemes are zeros. In fact, with limited transmit power, a stringent QoS requirement $P^{\max}_{\mathrm{e}}$ in constraint C2 is more difficult to satisfy leading to an infeasibility of the optimization problem in \eqref{eq:proposed_formulation_origion}.
With the increase of $P^{\max}_{\mathrm{e}}$, the proposed scheme is the first one that admits feasible solutions showing its superiority over other baseline schemes. In particular, due to the joint optimization of the precoder and IRS phase shifts, the proposed scheme can exploit the spatial degrees of freedom more efficiently than that of baseline scheme 2 to fulfill the SER constraint C2.
Furthermore, since baseline scheme 1 does not consider constraint C2, the SER of decoding IRS information approaches $0.5$ which is not suitable for most practical applications.
Additionally, by comparing the average sum-rate of the proposed scheme and the upper bound, it can be observed that there is a performance gap between the proposed scheme and the upper bound when $P^{\max}_{\mathrm{e}}$ is small. This is mainly because the IRS phase shifts are forced to align to the channels of the SU for satisfying the more stringent SER constraint. This leads to a weakened signal received at the PU.  However, unlike the proposed scheme, the upper bound cannot serve the primary and secondary system concurrently to realize a mutualistic SR system. Once constraint C2 becomes less stringent, the performance degradation of the proposed scheme is negligible compared with the upper bound, even though the proposed scheme turns off some of the IRS elements to convey IRS modulated information.
On the other hand, as $P^{\max}_{\mathrm{e}}$ further increases, the performance of all the schemes approaches a constant. This is because the limited transmit power budget $P_{\max}$ becomes the bottleneck of system performance, instead of $P_{e}^{\max}$.
In Fig. \ref{Fig:SumRVSP}, by further increasing the maximum transmit power budget at the AP, the system sum-rate increases monotonically. Indeed, by applying the proposed scheme, the AP and the IRS can effectively exploit the additional transmit power to create more powerful beamforming for improving the system sum-rate of the primary system. Moreover, the average sum-rates for both the proposed scheme and baseline scheme 2 grow as the number of the antennas at the AP increases due to an increasing beamforming gain. However, diminishing return appears when $N_{\mathrm{t}}$ is large as the result of channel hardening.
\vspace*{-3mm}
\section{Conclusion}\vspace*{-1mm}
In this paper, we proposed a MISO downlink SR communication system assisted by an IRS, which facilitates the primary transmission and mutualistic information transmission to the SU simultaneously. Different from existing works, we considered a more practical secondary system adopted a non-coherent detection at the SU. We derived an SER upper bound to characterize the non-coherent decoding performance. The joint design of the beamformer at the AP and the phase shifts at the IRS was formulated as a non-convex optimization problem to maximize the average system sum-rate taking into account the QoS requirement of decoding IRS symbols at the SU. Simulation results showed that the proposed scheme greatly enhances the performance of both the primary system and the secondary system significantly compared with some existing schemes.
\vspace*{-4mm}
\bibliography{IRSIT}

\begin{thebibliography}{10}
\providecommand{\url}[1]{#1}
\csname url@samestyle\endcsname
\providecommand{\newblock}{\relax}
\providecommand{\bibinfo}[2]{#2}
\providecommand{\BIBentrySTDinterwordspacing}{\spaceskip=0pt\relax}
\providecommand{\BIBentryALTinterwordstretchfactor}{4}
\providecommand{\BIBentryALTinterwordspacing}{\spaceskip=\fontdimen2\font plus
\BIBentryALTinterwordstretchfactor\fontdimen3\font minus
  \fontdimen4\font\relax}
\providecommand{\BIBforeignlanguage}[2]{{%
\expandafter\ifx\csname l@#1\endcsname\relax
\typeout{** WARNING: IEEEtran.bst: No hyphenation pattern has been}%
\typeout{** loaded for the language `#1'. Using the pattern for}%
\typeout{** the default language instead.}%
\else
\language=\csname l@#1\endcsname
\fi
#2}}
\providecommand{\BIBdecl}{\relax}
\BIBdecl

\bibitem{long2019symbiotic}
R.~Long, Y.-C. Liang, H.~Guo, G.~Yang, and R.~Zhang, ``Symbiotic radio: A new
  communication paradigm for passive {Internet of Things},'' \emph{IEEE
  Internet Things J.}, vol.~7, no.~2, pp. 1350--1363, Nov. 2019.

\bibitem{wu2019intelligent}
Q.~Wu and R.~Zhang, ``Intelligent reflecting surface enhanced wireless network
  via joint active and passive beamforming,'' \emph{IEEE Trans. Wireless
  Commun.}, vol.~18, no.~11, pp. 5394--5409, Aug. 2019.

\bibitem{yan2019passive}
W.~Yan, X.~Yuan, and X.~Kuai, ``Passive beamforming and information transfer
  via large intelligent surface,'' \emph{IEEE Wireless Commun. Lett.}, vol.~9,
  no.~4, pp. 533--537, Apr. 2020.

\bibitem{kammoun2020asymptotic}
A.~Kammoun, A.~Chaaban, M.~Debbah, M.-S. Alouini \emph{et~al.}, ``Asymptotic
  max-min {SINR} analysis of reconfigurable intelligent surface assisted {MISO}
  systems,'' \emph{IEEE Trans. Wireless Commun.}, vol.~19, no.~12, pp.
  7748--7764, Apr. 2020.

\bibitem{dong2020secure}
L.~Dong and H.-M. Wang, ``Secure {MIMO} transmission via intelligent reflecting
  surface,'' \emph{IEEE Wireless Commun. Lett.}, vol.~9, no.~6, pp. 787--790,
  Jan. 2020.

\bibitem{hua2021novel}
M.~Hua, Q.~Wu, L.~Yang, R.~Schober, and H.~V. Poor, ``A novel wireless
  communication paradigm for intelligent reflecting surface based symbiotic
  radio systems,'' \emph{arXiv preprint arXiv:2104.09161}, 2021.

\bibitem{zhang2021reconfigurable}
Q.~Zhang, Y.-C. Liang, and H.~V. Poor, ``Reconfigurable intelligent surface
  assisted {MIMO} symbiotic radio networks,'' \emph{IEEE Trans. Commun.},
  vol.~69, no.~7, pp. 4832--4846, Mar. 2021.

\bibitem{lin2020reconfigurable}
S.~Lin, B.~Zheng, G.~C. Alexandropoulos, M.~Wen, M.~D. Renzo, and F.~Chen,
  ``Reconfigurable intelligent surfaces with reflection pattern modulation:
  Beamforming design and performance analysis,'' \emph{IEEE Trans. Wireless
  Commun.}, vol.~20, no.~2, pp. 741--754, Feb. 2021.

\bibitem{liu2020deep}
C.~Liu, X.~Liu, D.~W.~K. Ng, and J.~Yuan, ``Deep residual learning for channel
  estimation in intelligent reflecting surface-assisted multi-user
  communications,'' \emph{IEEE Trans. Wireless Commun.}, Aug. 2021 [Early
  Access], doi: 10.1109/TWC.2021.3100148.

\bibitem{tse2005fundamentals}
D.~Tse and P.~Viswanath, \emph{Fundamentals of wireless communication}.\hskip
  1em plus 0.5em minus 0.4em\relax Cambridge university press, May 2005.

\bibitem{5453306}
H.-T. Wu, R.~Han, W.~Lerdsitsomboon, C.~Cao, and K.~K. O, ``Multi-level
  amplitude modulation of a 16.8-{GHz} class-{E} power amplifier with negative
  resistance enhanced power gain for 400-{Mbps} data transmission,'' \emph{IEEE
  J. Solid-State Circuits}, vol.~45, no.~5, pp. 1072--1079, Apr. 2010.

\bibitem{jeganathan2008spatial}
J.~Jeganathan, A.~Ghrayeb, and L.~Szczecinski, ``Spatial modulation: Optimal
  detection and performance analysis,'' \emph{IEEE Commun. Lett.}, vol.~12,
  no.~8, pp. 545--547, Aug. 2008.

\bibitem{jameson2016incomplete}
G.~Jameson, ``The incomplete {Gamma} functions,'' \emph{The Mathematical
  Gazette}, vol. 100, no. 548, pp. 298--306, Jul. 2016.

\bibitem{zhang2006schur}
F.~Zhang, \emph{The Schur complement and its applications}.\hskip 1em plus
  0.5em minus 0.4em\relax New York: Springer, 2005.

\bibitem{opial1967weak}
Z.~Opial, ``Weak convergence of the sequence of successive approximations for
  nonexpansive mappings,'' \emph{Bull. Am. Math. Soc.}, vol.~73, no.~4, pp.
  591--597, 1967.

\end{thebibliography}
\bibliographystyle{IEEEtran}

\end{document}